\documentclass[onecolumn,showpacs,superscriptaddress,prd]{revtex4}
\usepackage{amsmath}
\usepackage{amsfonts}
\usepackage{amssymb}
\usepackage[dvips]{graphicx}
\usepackage{hyperref}
\usepackage{units}
\usepackage{color}

\def\b{\begin{equation}}
\def\e{\end{equation}}
\def\ba{\begin{eqnarray}}
\def\ea{\end{eqnarray}}

\def\la{\langle}
\def\ra{\rangle}
\begin{document}

\title{Spacetime anisotropy affects cosmological entanglement}

\author{Roberto Pierini\footnote{email: roberto.pierini@unicam.it}}
\affiliation{School of Science and Technology, University of Camerino, I-62032 Camerino, Italy}
\affiliation{INFN-Sezione di Perugia, I-06123 Perugia, Italy}

\author{Shahpoor Moradi\footnote{moradis@ucalgary.ca}}
\affiliation{University of Calgary, Department of Geoscience, Calgary, Canada}

\author{Stefano Mancini\footnote{email: stefano.mancini@unicam.it}}
\affiliation{School of Science and Technology, University of Camerino, I-62032 Camerino, Italy}
\affiliation{INFN-Sezione di Perugia, I-06123 Perugia, Italy}

\date{\today}

\begin{abstract}
Most existing cosmological entanglement studies are focused on the isotropic Robertson-Walker (RW) spacetime. Here we go beyond this limitation and study the influence of anisotropy on entanglement generated by dynamical spacetime. Since the isotropic spacetime is viewed as a background medium and the anisotropy is incorporated as perturbation, we decompose entanglement entropy into isotropic and anisotropic contributions. The latter is shown to be non-negligible by analyzing two cosmological models with weak and conformal coupling. We also show the possibility of using entanglement to infer about universe features. 
\end{abstract}



\maketitle

\section{Introduction}

Relativistic quantum information, started out from the seminal paper \cite{MC}, is a new promising field that aims at understanding how special or general relativity can affect quantum information processing (see \cite{AF,RID,PT04,martin12,martin14} for reviews on different aspects of the theory). In particular, it has been understood that entanglement \cite{Horo}, a useful resource in quantum information science, is not a covariant quantity, which means that it depends on the observer \cite{Shi04,AFMT,FM}. Important examples can be found in cosmology, where the expansion of the universe can populate with particles and anti-particles the vacuum state \cite{LP}. The created pairs result entangled and such entanglement has been studied in \cite{ball06,fuentes10}, where it was shown that entanglement also stores information about the history of the universe (see also \cite{kanno15,steeg09,wang15}). As a further step, generation of entanglement between occupation numbers, vacuum and $1/2$-particle state, and polarizations (spins) of particles has been investigated \cite{MPM14,roberto16}. 
All such works employed Robertson-Walker (RW) models with the assumption of homogeneity and isotropy, so that it is yet unclear whether entanglement may be sensitive to spacetime anisotropies and to what extent universe's features are imprinted on it. Taking into account anisotropic models is necessary because of anisotropy found in the Cosmic Microwave Background radiation \cite{quercellini,weinberg}. Another open question is how emerging entanglement is affected when going beyond the conformal coupling case.

In the theory of quantum fields in curved spacetime (see e.g. \cite{BD01}), matter creation is associated with conformal symmetry breaking of the underling geometry, where the created pairs of particles and anti-particles with opposite momenta are entangled. We shall consider in the present paper three physical entities responsible for conformal symmetry breaking, namely mass of particles associated with the field, coupling between field and spacetime curvature, small gravitational disturbance (anisotropy).
We then characterize entanglement generated from the vacuum state by the anisotropic expansion of the universe assuming that we have access to particles modes only, and then evaluating the subsystem entropy. Given the relation between entanglement and cosmological parameters we show how it is possible to extract information regarding the underlying geometry from the detected entanglement. 

The paper is organized as follow. In Section \ref{section 2} we recall the Bogoliubov transformations and we show how to relate their coefficients to the amount of created entanglement. In Section \ref{section 3} we introduce a suitable cosmological model and explicitly compute the Bogolyubov coefficients for a scalar field. Next, in Section \ref{Ent}, we show how entanglement within this model is affected by mass of particles, field-spacetime coupling strength, and anisotropy. 
Section \ref{sec:para} deals with the possibility of deriving cosmological parameters from the detected entanglement.
The concluding section \ref{sec:conclu} is devoted to summarize the results and to discuss possible extensions of the present work. 
In Appendix \ref{app}, for the sake of comparison with Section \ref{Ent}, are shown results concerning a cosmological model with a different scale factor in the case of conformal coupling.


\section{Particles production and entanglement}\label{section 2}

Consider a scalar field $\phi(\vec x,t)$ and two spacetime regions, called $in$-region and $out$-region, which we assume to be Minkowskian. Let us define two different basis $\{\phi^{in}_i,\phi_i^{in*}\}$ and $\{\phi_i^{out},\phi_i^{out*}\}$, with support on the relative region only, which can be used to decompose the field. Elements in the basis set are orthonormal, which means $(\phi_i,\phi_j)=\delta_{ij}$, $(\phi_i^*,\phi_j^*)=-\delta_{ji}$ and $(\phi_i,\phi_j^*)=0$, according to the Klein-Gordon inner product, defined by 
\b\label{KG-inn}
(\phi_1,\phi_2)=-i\int_{\Sigma} d\Sigma^{\mu}\,\phi_1\overleftrightarrow{\partial_{\mu}}\phi_2^* \,,
\e
where $\phi_1\overleftrightarrow\partial_{\mu}\phi_2=\phi_1\partial_{\mu}\phi_2-\phi_2\partial_{\mu}\phi_1$ and $d\Sigma^{\mu}=n^{\mu}d\Sigma$, being $n^{\mu}$ a unit vector orthogonal to the spacelike hypersurface 
$\Sigma$ and $d\Sigma$ the volume element in $\Sigma$. Notice that the inner product (\ref{KG-inn}) is independent of $\Sigma$ \cite{BD01}.
After that, we can associate to each mode $\phi^{in/out}_i$ and to its complex conjugate $\phi_i^{in/out*}$ annihilation operators $ a_k^{in/out}$ and creation operators $a^{in/out\,\dagger}_{k}$ , such that they satisfy the usual commutation relations
$[ a_k^{in/out}\,, a^{in/out\,\dagger}_{k'}]=\delta_{kk'}$ and $[ a_k^{in/out}\,, a^{in/out}_{k'}]=0$.
Therefore, we have two equivalent representations of the scalar field
\begin{equation}\label{scadec}
\begin{aligned}
\Phi&=\sum_k\left\{ a^{in}_k\phi^{in}_k+a^{in\,\dagger}_k\phi^{in*}_k\right\}\\
&=\sum_{k'}\left\{a_{k'}^{out}\phi^{out}_{k'}+a^{out\,\dagger}_{k'}\phi_k^{out*}\right\}  \,.
\end{aligned}
\end{equation}
Each mode can now be expanded in terms of the others
\begin{equation}
\phi^{out}_{k'}=\sum_k\{ \alpha_{k'k}\phi^{in}_{k}+\beta_{k'k}\phi^{in*}_{k}\} \,,
\end{equation}
which, when substituted in (\ref{scadec}) implies the so-called Bogoliubov transformation and its inverse
\begin{equation}\label{dir-bog-tra} 
\begin{aligned}
a_k^{in}&=\sum_{k'}\{ \alpha_{k'k}a^{out}_{k'}+\beta_{k'k}^{*} a^{\dagger\,out}_{k'}\} \,,  \\
a^{out}_{k'}&=\sum_k\{ \alpha^*_{k'k} a^{in}_{k}-\beta^*_{k'k}a^{\dagger\,in}_{k}\}\,.
\end{aligned}
\end{equation}
These relate the ladder operators in the two different representations.
The coefficients $\alpha$ and $\beta$ are the well known Bogoliubov coefficients, defined as $\alpha_{ij}=(\phi^{out}_i,\phi_j^{in})$ and $\beta_{ij}=-(\phi^{out}_i,\phi_j^{in*})$.

Imposing now that $a^{\dagger\,out}_{k}$ and  $a^{out}_{k}$, defined through Eq.(\ref{dir-bog-tra}), satisfy the standard commutation relations, we get
\begin{equation}\label{bog-relations}
\begin{aligned}
\left[a^{out}_{k_1}, a^{\dagger\,out}_{k_2} \right]&=\sum_k\left( \alpha_{k_1k}^*\alpha_{k_2k}-\beta^*_{k_1k}\beta_{k_2k}  \right)=\delta_{k_1\,k_2},\\
\left[a^{\dagger\,out}_{k_1}, a^{\dagger\,out}_{k_2} \right]&=\sum_k\left( \alpha_{k_1k}\beta_{k_2k}-\beta_{k_1k}\alpha_{k_2k}  \right)=0   \,.
\end{aligned}
\end{equation}
We will be working with an homogeneous metric, meaning that the propagation of the field through the space can be decomposed into modes that will remain plane waves throughout, each with proper wave lenght, $\phi_k(x)=\chi_k(t)e^{i\vec k\cdot\vec x}$. Therefore, the inner product between modes gives a Dirac delta and the Bogoliubov coefficients have to be
\begin{equation}
\alpha_{k'k}=\alpha_k \,\delta_{k\,k'}   \qquad\qquad  \beta_{k'k}=\beta_{k} \,\delta_{k\,-k'} \,.
\end{equation}
Plugging these into (\ref{dir-bog-tra}), gives diagonal Bogoliubov transformations
\begin{equation}\label{inv-bog-tra1}
\begin{aligned}
a^{in}_k&=\alpha_{k} a^{out}_k+\beta_{k}^* a_{-k}^{\dagger\,out} \,,  \\
a^{out}_k&=\alpha^*_{k} a^{in}_k-\beta_{k}^* a_{-k}^{\dagger\,in} \,,
\end{aligned}
\end{equation}
whose main feature is that they do not mix modes.
Furthermore Eq.(\ref{bog-relations}) gives
\b\label{norm1}
|\alpha_k|^2-|\beta_k|^2=1   \,.
\e
The expectation value of the number operator for the $out$ mode, $N_k^{out}=a_{k}^{\dagger\,out}a_{k}^{out}$, in the vacuum state of the $in$ mode reads
\b\label{part-number}
{}_{in}\langle0|N^{out}|0\rangle_{in}=|\beta_{k}|^2  \,,
\e
thanks to Eq.(\ref{inv-bog-tra1}).  

In the context of cosmology, $|\beta_{k}|^2$ is the number of created particles per mode due to the dynamical nature of spacetime.
Let us consider now a vacuum state as seen by an observer in the early past ($in$ region) of our expanding universe. We focus only on modes $k$ for particles and $-k$ for antiparticles, so that the input state we are considering is $|0\ra_k^{in}|0\ra_{-k}^{in}$. Being a pure state of a bi-partite system, the $in$-vacuum can be seen by an observer in the far future ($out$ region) as
\b\label{sch-dec}
|0\ra_k^{in}|0\ra_{-k}^{in}=\sum_{n=0}^{\infty}c_n \;|n\ra_k^{out}|n\ra_{-k}^{out} \,.
\e
by using the Schmidt decomposition.
To find the coefficients $c_n$ of (\ref{sch-dec}), following \cite{ball06}, we use linear independence and the fact that $a_k|vac\ra=0$, obtaining
\b
c_n=\left(-\frac{\beta_{k}^*}{\alpha_k}\right)^n \, c_0 \,.
\e
Then, imposing normalization on (\ref{sch-dec}) we find that $c_0=\sqrt{1-\gamma}$, with 
\b\label{gamma}
\gamma=\left|\frac{\beta_{k}^*}{\alpha_k}\right|^2   \,.
\e
Note that $\gamma$ is always less than one because of (\ref{norm1}). What we want now is to measure the entanglement between particles and anti-particles modes in the physical state (\ref{sch-dec}). To accomplish it we consider, first of all, the corresponding density operator 
$\rho^{out}_{k\,,-k}$. After that, the amount of entanglement can be evaluated using the Von Neumann entropy of the reduced density operator (e.g. for particles) \cite{Horo} 
\b
S(k)=-Tr\left(\rho^{out}_k\log_2 \rho^{out}_k\right),
\e
where
\b\label{rho-k}
\rho^{out}_k=Tr_{-k}\,\left(\rho^{out}_{k,-k}\right)=(1-\gamma)\sum_{n=0}^{\infty}\gamma^n\,|n\ra_k\la n|^{out} \,.
\e
It then results
\b\label{sca-entr-funct}
S(k)=-\log_2(1-\gamma)-\frac{\gamma}{1-\gamma}\log_2\gamma=\log_2\left(\frac{\gamma^{\frac{\gamma}{\gamma-1}}}{1-\gamma}\right)\,.
\e


\section{Anisotropic cosmological model}\label{section 3}

Let us now look into a specific model of an anisotropic universe filled with a matter field.
The conformal symmetry breaking that leads to particles production can arise by the departure of the background spacetime from conformal flatness. Following \cite{BD02,ZS01}, let us consider a Bianchi $I$ spacetime with metric
\b
ds^2=dt^2-\sum_ia_i^2(t)dx_i^2 \,.
\e
Since we assume weak anisotropy, we write
$a_i(t)=1+h_i(t)$. Furthermore we introduce the time parameter $\eta=\int^t\frac{dt'}{a(t')}$ to get
\b\label{metric}
ds^2=a^2(\eta)\left\{d\eta^2-\sum_i\left[1+h_i(\eta)\right]\left(dx^i\right)^2\right\} \,,
\e
with $i=1,2,3$. The time parameter $\eta$ reduces to the so-called conformal time in the isotropic limit.
The perturbation is considered to be small, that is $\max |h_i(\eta)|\ll 1$ and also, to greatly simplify calculations, we require that $\sum_ih_i(\eta)=0$.
The following choice satisfies such conditions
\b\label{pert-func}
h_i(\eta)=e^{-\rho\,\eta^2}\cos{(\varepsilon\,\eta^2+\delta_i)} \,,
\e
where $\delta_i=\frac{\pi}{2},\frac{\pi}{2}+\frac{2\pi}{3},\frac{\pi}{2}+\frac{4\pi}{3}$. The cosmological parameters $\rho$ and $\varepsilon$ represent, respectively, the expansion rate of the universe and the oscillation frequency of the perturbation.
Let us now consider a scale factor which allows us to compute the Bogoliubov coefficients for a scalar field coupled with the spacetime just defined. 
That is
\b\label{con-sca-fac}
a(\eta)=1-\frac{\rho^2}{2(\rho^2+\eta^2)}  \,,
\e
which represents a contracting universe bouncing back at $\eta=0$ and expanding out again \cite{BD02}.
In the asymptotic limit, $a(\eta\to\pm\infty)$ is equal to one so the spacetime becomes static and flat. In this model the Ricci curvature scalar takes the form \cite{ BD02}
\b\label{ricci-scalar}
R(\eta)=6\rho^2\frac{\eta^2-\rho^2}{(\eta^2+\rho^2/2)^3} \,.
\e

\subsection{Scalar field}

A massive bosonic field of mass $m$ obeys the Klein-Gordon equation
\b\label{KG}
\left(\Box_g+m^2+\xi R(\eta)\right)\phi(\vec x,\eta)=0 \,,
\e
where $\Box_g$ is the D'Alambertian generalized to the metric $g$ and the factor $\xi$ represents the coupling constant of the field with the Ricci scalar curvature $R$.
Since the spacetime is homogeneous, the solution of the field equation \eqref{KG} can be separated as
\b\label{eig-sol}
\phi_k(\vec x,\eta)=\frac{1}{(2\pi)^{3/2}}\,\frac{1}{a(\eta)}e^{i\vec k\cdot\vec x}\chi_k(\eta) \,.
\e
The function of the time parameter, $\chi_k(\eta)$, satisfies
\b\label{time-eq}
\ddot\chi_k(\eta)+[\omega_k^2-V_k(\eta)]\chi_k(\eta)=0 \,,
\e
where
$\omega^2=k^2+m^2a^2(\infty)$ and
\b\label{V-func}
V_k(\eta)=\sum_ih_i(\eta)k_i^2+m^2\left[a^2(\infty)-a^2(\eta)\right]-\Lambda R(\eta)a^2(\eta)\,,
\e
with $\Lambda=\xi-1/6$.
The function $V_k(\eta)$ is the sum of three contributions
\begin{align}\label{Vs}
&V_k^{(mass)}(\eta)=m^2\left[a^2(\infty)-a^2(\eta)\right]\,, \nonumber\\
&V_k^{(coup)}(\eta)=-\Lambda R(\eta)a^2(\eta) \,,\\
&V_k^{(aniso)}(\eta)=\sum_ih_i(\eta)k_i^2 \,, \nonumber
\end{align}
coming from mass, curvature coupling and anisotropy respectively. Two regimes are of special interest in cosmology: one is when $\xi=0$, called weak coupling, and the other is when $\xi=1/6$, called conformal coupling, for which $V^{(coup)}=0$.
We can think to the problem as in scattering theory, which means assuming the interaction of the gravitational field with the matter field to be zero in the early past and in the far future, i.e. $V_k(\pm\infty)=0$. Our choice of the involved functions is such that this condition is matched for every single contribution to the time dependent function $V_k(\eta)$, namely $\lim_{\eta\to\pm\infty}V_k^{(i)}=0$, with $i=\{(mass),(coup),(aniso)\}$.
The normalized free-wave solution, propagating from $\eta\to-\infty$, reads
\b\label{free-wave}
\chi^{in}_k(\eta)=\frac{1}{\sqrt{2\omega}}e^{-i\omega\eta} \,.
\e
The integral form of the differential equation (\ref{time-eq}) then becomes
\b\label{sol}
\chi_k(\eta)=\chi^{in}_k(\eta)+\frac{1}{\omega}\int_{-\infty}^{\eta}\,d\eta_1\,V_k(\eta_1)\sin\left(\omega(\eta-\eta_1)\right)\chi_k(\eta_1) \,.
\e
In late time regions $(\eta\to\infty)$, again, we have a free wave propagating
\b
\chi^{out}_k(\eta)=\alpha_{k}\chi^{in}_k(\eta)+\beta_{k}\chi_k^{in*}(\eta) \,,
\e
where the Bogolyubov coefficients $\alpha$ and $\beta$ results as
\begin{align}\label{bog-coef}
\alpha_{k}&=1+i\int_{-\infty}^{\infty}\chi_k^{in*}(\eta)V_k(\eta)\chi_k(\eta)\,d\eta \,, \\
\beta_{k}&=-i\int_{-\infty}^{\infty}\chi_k^{in}(\eta)V_k(\eta)\chi_k(\eta)\,d\eta \,.
\end{align}
Computing the Wronskian of the differential equation (\ref{sol}) results in equation (\ref{norm1}).
To solve equation (\ref{sol}) we resort to an iterative procedure. To the lowest order we have
\b\label{appr}
\chi_k(\eta)=\chi_k^{in}(\eta),
\e
and the Bogoliubov coefficients become
\begin{equation}\label{bog-coef}
\begin{aligned}
\alpha_{k}&=1+\frac{i}{2\omega}\int_{-\infty}^{\infty}V_k(\eta)\,d\eta \,,\\
\beta_{k}&=-\frac{i}{2\omega}\int_{-\infty}^{\infty}e^{-2i\omega\eta}V_k(\eta)\,d\eta \,.
\end{aligned}
\end{equation}
Inserting here the functions (\ref{Vs}) with the explicit form of $h_j(\eta)$ and $a(\eta)$ we can write the Bogoliubov coefficients as
\begin{equation}\label{bog-coef3}
\begin{aligned}
\alpha_{k}(\eta)&=1+\alpha^{(mass)}_{k}+\alpha^{(coup)}_{k}+\alpha^{(aniso)}_{k} \,  \\
\beta_{k}(\eta)&=\beta^{(mass)}_{k}+\beta^{(coup)}_{k}+\beta^{(aniso)}_{k}  \,,
\end{aligned}
\end{equation}
where
\begin{equation}
\begin{aligned}
\label{Bog-coef-expl}
&\alpha_k^{(mass)}=i\frac{7m^2\rho\pi}{16\omega} \,,\\
&\beta_k^{(mass)}=i\frac{m^2\pi\,\rho}{16\omega}(2\rho\omega-7)e^{-2\omega\rho}\,,\\
&\alpha_k^{(coup)}=i\frac{6\,\pi\Lambda}{\omega\rho}(5\sqrt{2}-7)\,,\\
&\beta_k^{(coup)}=i\frac{\pi\Lambda}{16\omega\rho}[(384\rho\omega+672)e^{-2\omega\rho}-480\sqrt{2}e^{-\sqrt 2\omega\rho}]  \,,\\
&\alpha^{(aniso)}_{k}= \frac{i\sqrt{\pi}}{2\omega}\sum_{j=1}^3 k_j^2 \,{\rm Re}\left\{ \frac{e^{-i\delta_j}}{\sqrt{\rho+i\varepsilon}} \right\}\,,\\
&\beta^{(aniso)}_{k}=- \frac{i\sqrt{\pi}}{2\omega}\sum_{j=1}^3 k_j^2 \,{\rm Re}\left\{ \frac{e^{-i\delta_j}\,e^{-\omega^2/(\rho+i\varepsilon)}}{\sqrt{\rho+i\varepsilon}} \right\}\,.
\end{aligned}
\end{equation}
We may notice that the Boguliubov coefficients (\ref{bog-coef3}) have a direct dependence on the mass $m$, on the coupling constant $\Lambda$ and on the anisotropy $h_i$. Each of these gives independent contributions to particles creation, through Eq.(\ref{part-number}), and to entanglement, through  Eq.(\ref{sca-entr-funct}).

Finally, we remark that caution must be used with the approximation of (\ref{sol}). Actually the normalization condition (\ref{norm1}) can be employed as a test for its validity \cite{Bir01}. In what follows, violations of (\ref{norm1}) within one-percent will be considered acceptable.


\section{Entanglement}\label{Ent}

Once we know the Bogoliubov coefficients for a specific model, we can use them to evaluate the
particle-antiparticle entanglement according to Eq.(\ref{sca-entr-funct}). In this section we will consider the subsystem entropy in the two relevant cases of weak coupling ($\xi=0$) and conformal coupling ($\xi=1/6$). To realize two dimensional plots we have made the ansatz
\b
k_i=\frac{k}{\sqrt{2+b}}\left\{1,1,\sqrt{b}\right\} \,,
\e
such that $ k=\sqrt{k_1^2+k_2^2+k_3^2}$. The amount of anisotropy is therefore quantified by the parameter $b$. If $b=1$ the anisotropy contribution in (\ref{V-func}) would disappears because of the condition $\sum_i h_i(\eta)=0$.

\subsection{Weak coupling}

\begin{figure}[ptb]
\includegraphics[width=2.5in]{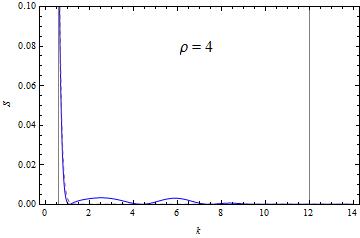}
\includegraphics[width=2.5in]{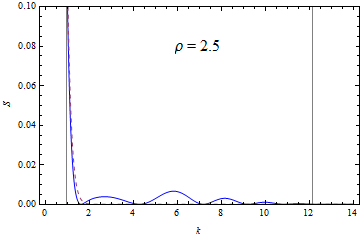} \\
\includegraphics[width=2.5in]{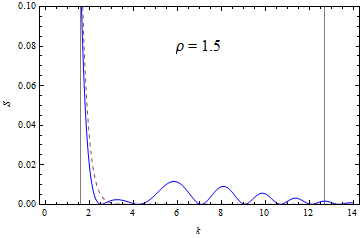}
\includegraphics[width=2.5in]{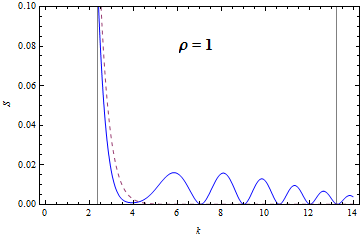}
\caption{
Isotropic (dashed line) and total subsystem entropy (solid line) $S$ vs $k$ for the scalar field, with with various values of $\rho$.
Here $b=1.1$, $\varepsilon=10$, $\xi=0$ and $m=0.01$.}
\label{fig1}
\end{figure}

\begin{figure}[ptb]
\includegraphics[width=2.5in]{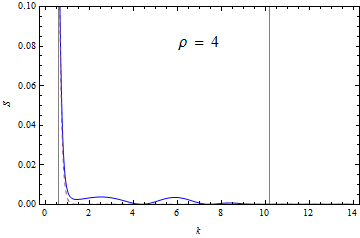}
\includegraphics[width=2.5in]{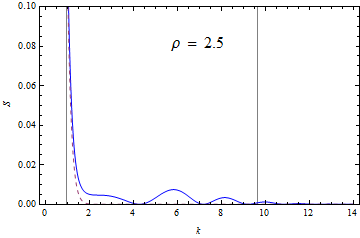}\\
\includegraphics[width=2.5in]{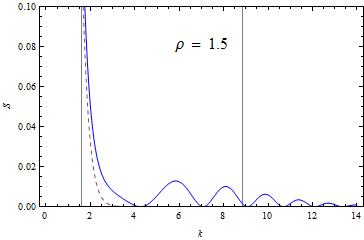} 
\includegraphics[width=2.5in]{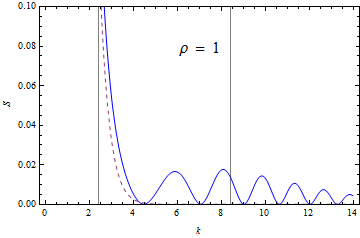}
\caption{
Isotropic (dashed line) and total subsystem entropy (solid line) $S$ vs $k$ for the scalar field, with various values of $\rho$.
Here $b=0.9$, $\varepsilon=10$, $\xi=0$ and $m=0.01$.}
\label{fig2}
\end{figure}

The total entanglement entropy and the isotropic entanglement entropy (where for isostropic we mean the contribution from mass and curvature coupling) have been plotted in Figures \ref{fig1} and \ref{fig2}, for $b>1$ and $b<1$ respectively, and with $\xi=0$.
The validity interval of the approximation (\ref{appr}) is delimited by two vertical lines. This is sensitive to cosmological parameters and therefore to the choice of the relevant functions (\ref{pert-func}),(\ref{con-sca-fac}). 

The problem is twofold: for large $k$ the anisotropic coefficients clearly are not the right ones, while for small $k$ the isotropic part is troublesome. In fact when the mass is large, the approximation works well for the coupling term and badly for the mass term, while the opposite occurs for small mass.
Anyway, even if our results are trustworthy only in a limited range of $k$, which generally exludes zero, very interesting features show up. The main is the fact that the total entropy oscillates as a function of $k$ after the contribution from isotropic quantities is dropped to zero. This means that for momenta higher than a certain value anisotropy plays a decisive role in creating and driving entanglement. Also, as expected, the oscillations are damped and their amplitude increase as $\rho$ decreases. As $\rho$ increases the entanglement entropy shrinks towards small momenta and oscillations tend to disappear. 
In this regime the isotropic contribution is dominant.
When $b>1$ the first oscillation due to anisotropy always starts at zero entropy, because the total entropy drops to zero faster than the isotropic one, while this is not the case for $b<1$, see Fig.\ref{fig2}.

\begin{figure}[ptb]
\includegraphics[width=2.5in]{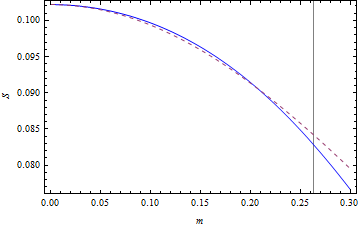}
\includegraphics[width=2.5in]{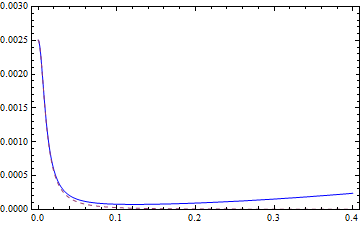}
\caption{
Left - Isotropic (solid line) and coupling (dashed line) subsystem entropy $S$ vs $m$ for the scalar field, with $k=1$, $\rho=2.5$ and $\xi=0$.  
Right - Total (solid line) and mass (dashed line) subsystem entropy $S$ vs $k$ for the scalar field, with 
with $m=0.01$, $\rho=1$ and $\xi=1/6$.}
\label{fig3}
\end{figure}

Let us have a closer look at the relation between mass and curvature coupling, neglecting for a while anisotropy. 
In Fig.\ref{fig3}-Left it can be seen the isotropic entropy and the mass entropy vs $m$. Although the approximation made to find the solution fails quite soon, we can argue that as the mass grows from zero its contribution becomes more significative, because the separation between the two curves increases. The curves are decreasing as a consequence of the dependence of $\beta_k$s on the negative exponents of the frequency $\omega$ (\ref{Bog-coef-expl}), which is an increasing function of the mass $m$. Particles (hence entanglement) creation are in general inhibited for high energetic particles. Also, the contribution from the mass is expected to be more important for heavy particles, but in our model this regime 
cannot be explored. As already mentioned, the approximation (\ref{app}) works well at low momenta for the mass contribution when particles are light, while for the coupling contribution it works well when particles are heavy.
 In general we can say that the contribution to the entanglement from coupling to curvature is monotonically decreasing vs $\xi$, with maximum at minimal coupling $\xi=0$.

To give a quantitative estimate of the relevance of anisotropy on entanglement, we can consider the ratio 
$\mathcal{R}$ for the total entropy between maximal oscillations' amplitude and maximum value taken at the lowest value of $k$ admitted by our approximation. It results that $\mathcal{R}$ increases from $0.025$ when $\rho=4$, up to $0.136$ when $\rho=1$. This means that anisotropic effects are non negligible, unless $\rho$ gets very large.

\subsection{Conformal coupling}

\begin{figure}[ptb]
\includegraphics[width=2.5in]{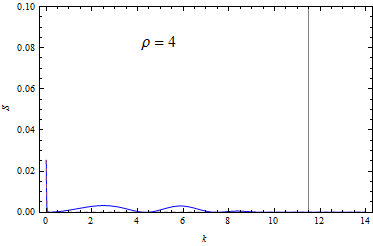}
\includegraphics[width=2.5in]{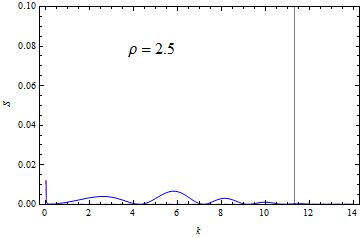} \\
\includegraphics[width=2.5in]{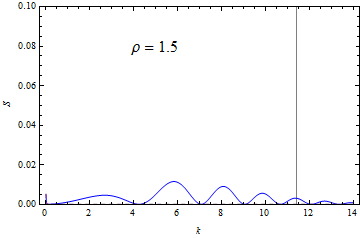}
\includegraphics[width=2.5in]{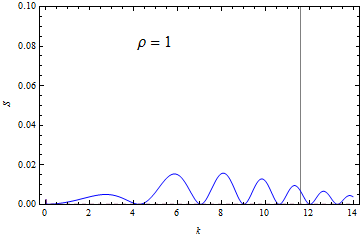}
\caption{
Isotropic (dashed line) and total subsystem entropy (solid line) $S$ vs $k$ for the scalar field, with various values of $\rho$.
Here $b=1.1$, $\varepsilon=10$, $\xi=1/6$ and $m=0.01$.}
\label{fig6}
\end{figure}

In the case of conformal coupling, $\xi=1/6$, the contribution to the isotropic part of the entanglement entropy comes from mass only. A relevant difference with the previous case is that now the normalization condition (\ref{norm1}) holds until $k=0$ for the range of $\rho$ we have considered in the previous subsection and for $m=0.01$, see Fig.\ref{fig3}-Right and Fig.\ref{fig6}. For larger values of $m$ and of $\rho$ deviations start at smaller values of $k$. Also, as $\rho$ decreases the maximum in Fig. \ref{fig6} decreases, in agreement to what found in Ref.\cite{fuentes10}. There, it was shown that an asymptotic regime is reached by the entropy when $\rho\to\infty$. Unfortunately, we are not able to study this regime here because of the sensibility of our approximation to cosmological parameters.
The important thing to note in Fig.\ref{fig3}-Right is that the total entropy and mass entropy completely overlap for small values of $k$, meaning that in such a regime anisotropy is completely negligible. Furthermore, since the dashed line in Fig.\ref{fig3} and \ref{fig6} drops to zero soon, we can ascribe the contribution of isotropic entropy in the Figures \ref{fig1} and \ref{fig2} to non-conformal coupling and conclude that the last is the most significative among the three contributions when $\xi=0$. 

\begin{figure}[ptb]
\includegraphics[width=2.5in]{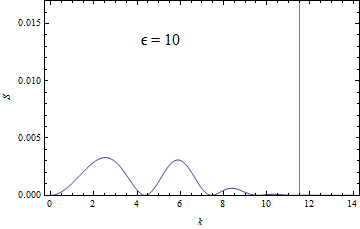}
\includegraphics[width=2.5in]{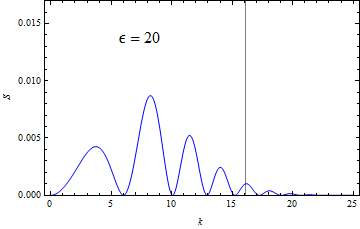}  \\
\includegraphics[width=2.5in]{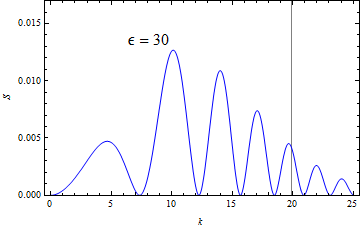}
\includegraphics[width=2.5in]{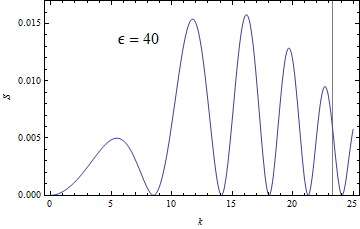}
\caption{Anisotropic subsystem entropy $S$ vs $k$ for the scalar field, with various values of $\varepsilon$. Here $b=1.1$ and $\rho=4$.}
\label{fig5}
\end{figure}

In Fig.\ref{fig5} we show the solely anisotropic part of the entanglement entropy, corresponding to a massless field conformally coupled with gravity. It is clearly shown that the amplitude of the oscillations increases with the cosmological parameter $\varepsilon$ and they shift towards higher momenta. 
For conformal coupling the ratio $\mathcal{R}$ turns out to be increasing from $0.130$ when $\rho=4$, up to $6.298$ when $\rho=1$. Here, the value of the entanglement entropy at the lowest acceptable $k$ (which is $k=0$ for the conformal coupling), becomes lower than the maximal oscillations' amplitude as $\rho$ decreases and hence the ratio $\mathcal{R}$ becomes greater than one.

The case of conformal coupling has been also considered in Appendix \ref{app}, but with a different scale factor.


\section{Cosmological parameters from entanglement}\label{sec:para}

In this Section we show how to extract information about the cosmological model from entanglement. Following the idea of Ref.\cite{ball06} this amounts to express cosmological parameters in terms of entanglement entropy. First we note that being the entanglement for scalar field (\ref{sca-entr-funct}) a monotonic growing function of $\gamma$, if we know $S(\gamma)$ we can consequently invert it and uniquely determine $\gamma$. 
Then, from the definition (\ref{bog-coef3}) and the condition (\ref{norm1}) it is possible to write
\b\label{ent-contributions}
|\alpha^{(coup)}|+|\alpha^{(mass)}|+|\alpha^{(aniso)}|=\sqrt{\frac{\gamma}{1-\gamma}}  \,,
\e
where the functions on the left hand side of the equation are given in (\ref{Bog-coef-expl}).
Let us now assume that the momentum of the detected particle is in the region were only the anisotropic coefficients are important (we know from Section \ref{Ent} that this region exists indeed). Thus, we can set to zero $\alpha^{(coup)}$ and $\alpha^{(mass)}$ in (\ref{ent-contributions}). The relevant quantities are now the parameters $b$, which somehow quantifies the anisotropy, and $\varepsilon$, the oscillation frequency. In our simple model where $k_i=\frac{k^2}{\sqrt{2+b}}\{ 1,1,\sqrt{b} \}$, the parameter $b$ can be expressed as
\b\label{b-par}
b=\left\{
\begin{array}{ccc}
\frac{4\omega\sqrt{r}\sqrt{\gamma/(1-\gamma)}-k^2\sqrt\pi(\lambda_{1}+\lambda_{2})}{k^2\sqrt\pi\lambda_3-2\omega\sqrt{r}\sqrt{\gamma/(1-\gamma)}}\,, &  & \lambda_1+\lambda_2+b\lambda_3\geq 0
\\
\\
-\frac{4\omega\sqrt{r}\sqrt{\gamma/(1-\gamma)}+k^2\sqrt\pi(\lambda_{1}+\lambda_{2})}{k^2\sqrt\pi\lambda_3+2\omega\sqrt{r}\sqrt{\gamma/(1-\gamma)}}\,, & & \lambda_1+\lambda_2+b\lambda_3<0
\end{array} 
\right.\,,
\e
where $\lambda_i=\cos(\delta_i+\phi/2)$, $r=\sqrt{\rho^2+\epsilon^2}$ and $\tan\phi=\frac{\varepsilon}{\rho}$.

To get $\varepsilon$ one has to solve the following fourth order equation
\begin{equation}
\begin{aligned}
4{\cal A}^2(\gamma,k)\varepsilon^4&+8\,{\cal A}(\gamma,k)\,\mathfrak{c}(b,k)\,\mathfrak{s}(b,k)\varepsilon^3
+\left[4{\cal A}^2(\gamma,k)\rho^2+4{\cal A}(\gamma,k)\rho\left(\mathfrak{s}^2(b,k)-\mathfrak{c}^2(b,k)\right)-\left(\mathfrak{s}^2(b,k)-\mathfrak{c}^2(b,k)\right)^2\right]\varepsilon^2  \nonumber\\
&+4\rho\,\mathfrak{c}(b,k)\,\mathfrak{s}(b,k)\left[\mathfrak{s}^2(b,k)-\mathfrak{c}^2(b,k)
+{\cal A}(\gamma,k)\rho\right]\varepsilon\nonumber \\
&=\rho^2\left(\mathfrak{c}^2(b,k)+\mathfrak{s}^2(b,k)\right)^2
-\left[\rho\left(\mathfrak{s}^2(b,k)-\mathfrak{c}^2(b,k)\right)+{\cal A}(\gamma,k)\rho^2\right]^2  \,,
\end{aligned}
\end{equation}
where ${\cal A}(\gamma,k)=\frac{4\omega^2}{\pi}\,\frac{\gamma}{1-\gamma}$, 
$\mathfrak{c}(b,k)=\sum_ik_i^2\cos\delta_i$ and $\mathfrak{s}(b,k)=\sum_ik_i^2\sin\delta_i$.

Much easier would be getting an expression for the parameter $\rho$, assuming the momentum of the detected particle is in the region were only the contribution from coupling with curvature matters. This region is just before the anisotropic region as seen in Section \ref{Ent}.
For massless particles, setting $\alpha^{(mass)}=0$ and $\alpha^{(aniso)}=0$, we can easily find from eq. (\ref{ent-contributions})
\b
\rho=6\left| (7-5\sqrt 2)\,\frac{\Lambda\pi}{\omega}\right| \,\sqrt{\frac{1-\gamma}{\gamma} } \,,
\e

If we assume $\rho\gg\varepsilon$ that would give $r\simeq\rho$ 
and $\cos(\delta_i+\phi/2)\sim\sqrt 2\,\cos\delta_i$. 
In contrast, if $\rho\simeq\varepsilon$ it gives $r\simeq\rho\sqrt 2$ and
$
\cos(\delta_i+\phi/2)\simeq\sqrt{\frac{\sqrt 2+1}{2\sqrt 2}}\cos\delta_i-\sqrt{\frac{\sqrt 2-1}{2\sqrt 2}}\sin\delta_i \nonumber\,.
$


\section{Concluding Remarks}\label{sec:conclu}

Summarizing, we have investigated particle-antiparticle entanglement arising from conformal symmetry breaking induced by three factors: mass of particles associated with the field, coupling between field and spacetime curvature, small gravitational disturbance (anisotropy). 
To this end we have viewed the isotropic spacetime as a background medium and considered the anisotropy as perturbation. We have then shown that, in the range of validity of the first order approximate solution of scalar field equation, anisotropy provides non negligible effects on the entanglement spectrum (entanglement entropy vs momentum). In particular it changes its behavior (introducing oscillations) and for some values of momentum it results the only contribution.
While conformal coupling does not give contribution to the entanglement spectrum, 
weak coupling gives the most relevant one.
The mass is significant only for very small value of momentum. 
This is also confirmed by the considerations in Appendix \ref{app} of a different scale factor, which additionally shows the possibility of entanglement generation even for a massless field.

The ranges of momentum where the various contributions are important are almost separated between isotropic and anisotropic part. This fact can be used to get a system of equation, relating entanglement and Bogoliubov coefficients, from which the relevant cosmological parameters can be explicitly computed.
Obviously this kind of calculation should be refined with more realistic cosmological models.

Entanglement has been studied in cosmological isotropic settings also for half spin particles \cite{fuentes10,MPM14,wang15,roberto16}. There, in particular, 
it was shown that the Dirac field is better suited than the scalar field to get information regarding the dynamics of the universe \cite{fuentes10}. Therefore it seems to be important looking at half spin particles in a more general background as we did for scalar particles in the present paper. This is left for future work.


\appendix
\section{Entanglement from a different scale factor}\label{app}

In Section \ref{Ent} we were able to show the behaviour of entanglement entropy for small values of $k$ only in the conformally-coupled case and for very low mass. Here, we will consider a different scale factor with respect to (\ref{con-sca-fac}) which gives us the opportunity to compare the anisotropic contribution with the isotropic one, when the mass contribution is relevant and the field is conformally coupled.

A contracting and expanding universe, with asymptotically Minkowskian regions, is described also by the following scale factor \cite{BD02}
\b\label{sc-fac}
a(\eta)=1-A e^{-\rho^2\eta^2} \,,
\e
where $\rho$ is the expansion rate and $A$ gives the minimum value of the scale factor $a(\eta=0)=1-A$.
The mass contribution to the Bogoliubov coefficients then becomes
\begin{align}\label{sca-coef}
&\alpha_{k}^{iso}=\frac{i m^2 A\sqrt{\pi}}{2\omega\rho}[2-A\frac{\sqrt 2}{2}],  \nonumber\\
&\beta^{iso}_{k}=-\frac{i m^2 A\sqrt{\pi}}{2\omega\rho}e^{-\omega^2/\rho^2}[2-A\frac{\sqrt 2}{2}e^{\omega^2/2\rho^2}],
\end{align}
while the anisotropic contribution is the same as in (\ref{Bog-coef-expl}).
Referring to the equation (\ref{sca-entr-funct}), using (\ref{sca-coef}) and the anisotropic coefficients in (\ref{Bog-coef-expl}), we get
\b\label{absganiso}
\gamma=\pi\,\frac{\left[m^2\,A\,(2-\frac{A}{\sqrt 2}e^{\omega^2/2\rho^2})e^{-(\omega^2/\rho^2)}+\frac{\rho}{\sqrt{r}}\,e^{-(\omega^2\rho/r^2)}\sum_ik_i^2\cos(\delta_i+\frac{\phi}{2}-
\frac{\varepsilon\omega^2}{r^2})\right]^2}
{4\omega^2\rho^2+\pi\left[m^2\,A(2-\frac{A}{\sqrt 2})+\frac{\rho}{\sqrt{r}}\sum_ik_i^2\cos(\delta_i+\frac{\phi}{2})\right]^2 }\,,
\e
where, like in Sec.\ref{sec:para}, we have $r=\sqrt{\rho^2+\varepsilon^2}$ and $\tan\phi=\frac{\varepsilon}{\rho}$.

\begin{figure}[ptb]
\includegraphics[width=2.5in]{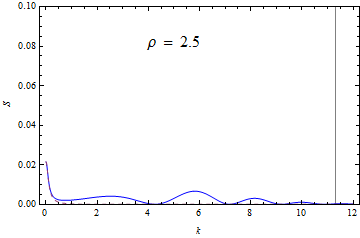}
\includegraphics[width=2.5in]{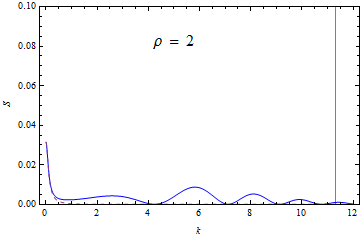} \\
\includegraphics[width=2.5in]{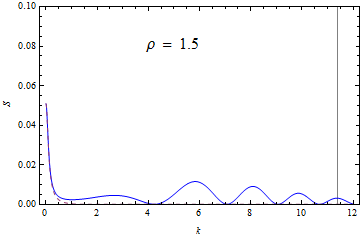}
\includegraphics[width=2.5in]{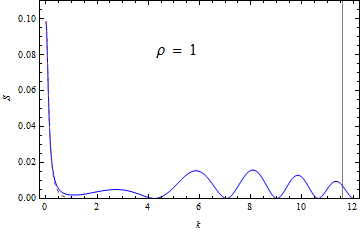}
\caption{Isotropic (dashed line) and total subsystem entropy $S$ vs $k$ for the scalar field, 
with various values of $\rho$.
Here $b=1.1$, $A=0.99$ and $m=0.1$.}

\label{fig7}
\end{figure}

\begin{figure}[ptb]
\includegraphics[width=2.5in]{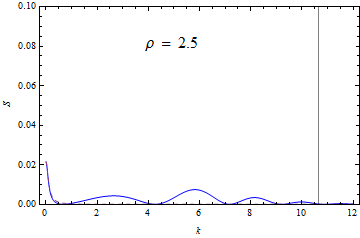}
\includegraphics[width=2.5in]{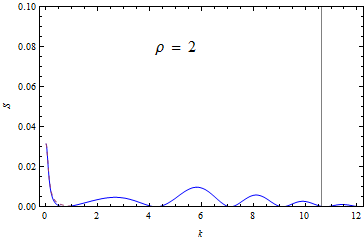}  \\
\includegraphics[width=2.5in]{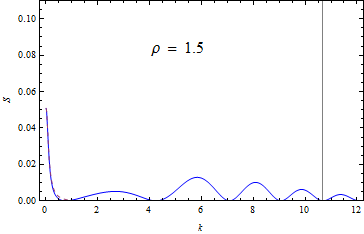}
\includegraphics[width=2.5in]{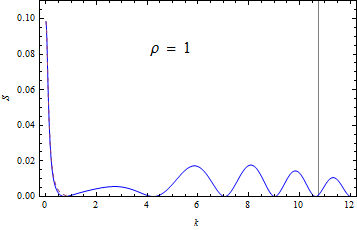}
\caption{Isotropic (dashed line) and total subsystem entropy $S$ vs $k$ for the scalar field, 
with with various values of $\rho$.
Here $b=0.9$, $A=0.99$ and $m=0.1$.}
\label{fig8}
\end{figure}

In Figures \ref{fig7} and \ref{fig8} we can see that the isotropic entropy starts from a maximum value greater than zero at $k=0$, and then drops to zero quite fast for increasing values of $k$, and subsequently oscillations start to show up. As the cosmological parameter $\rho$ decreases, the entropy peak at $k=0$ increases,
as opposite to the behaviour appearing with the scale factor (\ref{con-sca-fac}).
In the isotropic case, where $h_i=0$, expression (\ref{absganiso}) reduces to
\b\label{absgiso}
\gamma=\frac{\left[2 e^{-\omega^2/\rho^2}-\frac{A}{\sqrt 2}e^{-\omega^2/2\rho^2}\right]^2}{(2-\frac{A}{\sqrt 2})^2+\frac{4\omega^2\rho^2}{\pi m^4 A^2}} \,,
\e
which is a clear monotonic decreasing function of $k$ and its maximum, at $k=0$, depends on the value of the cosmological parameter $\rho$. Here, $\gamma\to 0$ as long as $m\to 0$, which means that entanglement is not present in the massless limit. This is expected because in this case there is no particles creation.
In contrast, taking $m\to 0$ when the conformal symmetry breaking - and the subsequent particles creation- is due to anisotropy of spacetime only, we have from (\ref{absganiso})
\b
\gamma=\pi\,\frac{\left[\sum_ik_i^2\cos\left(\varepsilon \,k^2/r^2-\delta_i-\frac{\phi}{2}\right)\right]^2}
{4k^2\,r+\pi\left[\sum_ik_i^2\cos\left(\delta_i+\frac{\phi}{2}\right)\right]^2 }\,e^{-2( k^2\rho/r^2)}\,.
\e
This case is particularly interesting because entanglement is generated even for a massless field.
Something that does not happen for isotropic spacetime.
The total entropy oscillates as $k$ gets larger than zero. Plots in Fig. \ref{fig7} clearly show that the first oscillation appears when the isotropic contribution is already zero, so that the oscillatory behavior has to be ascribed only to anisotropy. As $\rho$ decreases the maximum of anisotropic contribution does not change appreciably, showing a weak dependence on this cosmological parameter.
Notice also the difference in the behaviour of the total entropy when $b$ is lower or greater than zero. In the first case it falls down to the value zero faster than the isotropic curve, while in the second case slower.
The behaviour of the ratio $\mathcal{R}$ is now opposite to that of Sec.\ref{Ent}.B and we have 
$\mathcal{R}=0.309$ for $\rho=2.5$ decreasing up until $\mathcal{R}=0.160$ for $\rho=1$.
This means that anisotropic effects are non negligible, unless $\rho$ gets very small, in which situation by the way the approximation (\ref{appr}) is no longer valid.

\end{document}